\documentclass[prb,twocolumn,tbtags,superscriptaddress,floatfix]{revtex4}
\usepackage{amsmath}
\usepackage{graphicx}
\usepackage{amssymb}
\usepackage{color,soul}
\usepackage[T1]{fontenc}
\usepackage{ae,aecompl}
\usepackage{natbib}

\begin{document}

\title{Probabilistic motional averaging}

\begin{abstract}
	In a continuous measurement scheme a spin-1/2 particle can be measured and
	simultaneously driven by an external resonant signal. When the driving is weak, it does not prevent the particle wave-function from collapsing and a
	detector randomly outputs two responses corresponding to the states of the
	particle. In contrast, when driving is strong, the detector returns a single
	response corresponding to the mean of the two single-state responses. This
	situation is similar to a motional averaging, observed in nuclear magnetic
	resonance spectroscopy. We study such quantum system, being periodically
	driven and probed, which consists of a qubit coupled to a quantum resonator.
	It is demonstrated that the transmission through the resonator is defined by
	the interplay between driving strength, qubit dissipation, and resonator
	linewidth. We demonstrate that our experimental results are in good agreement with numerical and analytical calculations.
\end{abstract}

\author{D.~S.~Karpov}
\email{karpov@ilt.kharkov.ua}
\affiliation{B. Verkin Institute for Low Temperature Physics and Engineering, Kharkov,
	Ukraine}
\affiliation{Leibniz Institute of Photonic Technology, Jena, Germany}
\author{V.~Y.~Monarkha}
\affiliation{ARC Centre of Excellence for Engineered Quantum Systems, Queensland 4072,
	Australia}
\affiliation{Okinawa Institute of Science and Technology Graduate University, Onna,
	904-0495 Okinawa, Japan}
\author{D.~Szombati}
\affiliation{ARC Centre of Excellence for Engineered Quantum Systems, Queensland 4072,
	Australia}
\author{A.~G.~Frieiro}
\affiliation{ARC Centre of Excellence for Engineered Quantum Systems, Queensland 4072,
	Australia}
\affiliation{School of Mathematics and Physics, University of Queensland, St Lucia,
	Queensland 4072, Australia}
\author{A.~N.~Omelyanchouk}
\affiliation{B. Verkin Institute for Low Temperature Physics and Engineering, Kharkov,
	Ukraine}
\author{E.~Il'ichev}
\affiliation{Leibniz Institute of Photonic Technology, Jena, Germany}
\affiliation{Novosibirsk State Technical University, Novosibirsk, Russia}
\author{A.~Fedorov}
\affiliation{ARC Centre of Excellence for Engineered Quantum Systems, Queensland 4072,
	Australia}
\author{S.~N.~Shevchenko}
\affiliation{B. Verkin Institute for Low Temperature Physics and Engineering, Kharkov,
	Ukraine}
\affiliation{V. N. Karazin Kharkov National University, Kharkov, Ukraine}
\maketitle

\section{Introduction}

Circuit quantum electrodynamics studies scalable solid-state quantum
systems, behaving analogous to the interacting light and matter, with superconducting qubits playing the role of artificial atoms. At this point, a number of quantum
phenomena, known from atomic and optical physics, have been demonstrated in
solid-state systems, such as lasing and cooling of the electromagnetic field
in a resonator \cite{Astafiev07, Ashhab09, Grajcar08, Nori08, Hauss08, You08} and Mollow triplet \cite%
{Zhou08, Astafiev10, Greenberg17, Abdumalikov11}. Another phenomenon, called the 
\textit{motional averaging} effect, which originated from nuclear magnetic resonance spectroscopy \cite{Abragam}, has been also recently demonstrated
with a superconducting qubit~\cite{Li13}. In that experiment, the qubit was
driven by stochastic pulses. At the same time, a two tone spectroscopy was performed, which showed that the two spectroscopic lines (corresponding to two different states) are converted to a single broadened line for slow jumping rates, showing an increase of the average jumping rate~\cite{Li13, Abragam}. Different
aspects of the motional averaging were studied recently, such as classical
analogies~\cite{Ivakhnenko18}, weighted averaging~\cite{Ono19} and related
studies (see Refs.~\cite{Kono17, Pan17}).

\begin{figure}[ht]
\centering
\includegraphics[width=7cm]{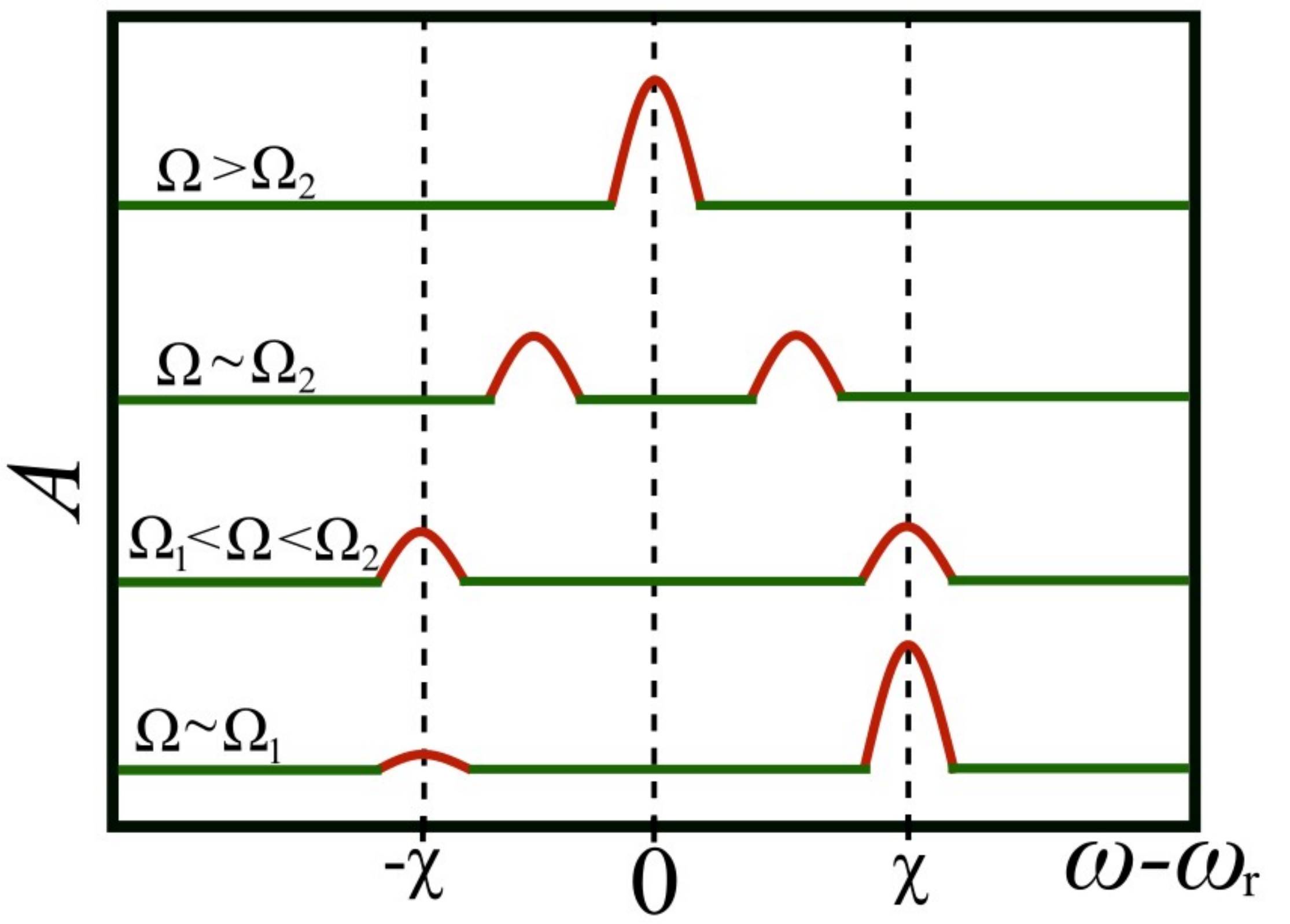}
\caption{Schematic of probabilistic motional averaging. If driven by a
periodic signal, qubit's response (the resonator transmission amplitude $A$)
is defined by the energy-level occupation probabilities. The height of the
two peaks at $\mathrm{\Omega} \sim \mathrm{\Omega} _{1}$ is defined by the upper-level
occupation probability and their position is defined by the dispersive shift 
$\protect\chi $. With increasing $\mathrm{\Omega} $, the two peaks, first, become of
equal height and then merge into one, which we term as a probabilistic
motional averaging. The characteristic frequency $\mathrm{\Omega} _{1}$ denotes when
the upper-level occupation probability becomes significant and $\mathrm{\Omega} _{2}$
is in-between the two regimes: (i) when the peaks are at $\protect\omega -%
\protect\omega _{\mathrm{r}}=\pm \protect\chi $ and (ii) when they merge
into one, at $\protect\omega =\protect\omega _{\mathrm{r}}$.}
\label{Fig1}
\end{figure}

Reminiscent of motional averaging can be observed in a qubit-resonator
system where, instead of a noisy signal, the qubit is driven by a periodic
pump, with frequency $\omega _{\mathrm{d}}$ and amplitude $\mathrm{\Omega}$. The
resonator which is dispersively coupled to the qubit, can be probed through
its own driving term with frequency $\omega $. The resulting transmission
through the resonator exhibits two resonant lines for the weak qubit driving
corresponding to the qubit states and a single line in between for when the
qubit driving is increased, see Fig.~\ref{Fig1}. This was demonstrated
recently for a transmon-based system and termed \textit{quantum rifling}~%
\cite{Szombati19} due to the lack of measurement back-action on the qubit
from the driven resonator to the strongly driven qubit; see also Refs.~\cite{Ashhab09b, Ashhab09c}. In this paper, we do
not study measurement back-action but focus on the motional averaging
picture for the resonator line and obtain simplified analytical expressions
for resonator transmission.

Interaction of a quantum system with a quantized electromagnetic field is
considered in the frame of the quantum Rabi model~\cite{Schleich,
ScullyZubairy, Gu17, Kockum19, Pirkkalainen14, Silveri13, Saiko17-1}. In the situation where the coupling strength $\mathrm{g}$
between the two-level system and the electromagnetic field is much smaller
than the difference between the qubit $\omega _{\mathrm{q}}$ and resonator $%
\omega _{\mathrm{r}}$ frequencies, $\left\vert \omega _{\mathrm{q}}-\omega _{%
\mathrm{r}}\right\vert >\mathrm{g}$, this is known as the Jaynes-Cummings
model~\cite{ScullyZubairy, Zagoskin2011, Wendin16, Shevchenko, Saiko14}.

Experiments are largely described by the semi-classical approximation, e.g.
Refs.~%
\cite{Astafiev07, You07b, Grajcar08, Hauss08, Scarlino18, Silveri15, Ian10, Oelsner13, Shevchenko14, Karpov16,
Oelsner18, Chang19}, when a chain of equations is restrained by keeping
only the first-order correlators. To describe time-evolution experiments, a semi-quantum approach is more correct~\cite{Szombati19,
Bianchetti09, Andre09}; comparison of the two approaches can be found e.g.
in Ref.~\cite{Shevchenko18}. Note that in practice, to get a
spectrum of a non-linear level structure of the system, one should apply two
signals, to drive the system and a weak probe tone, which is known as a
two-tone spectroscopy~\cite{Saiko17, Oelsner18, Kohler18}.

In this paper, we study measurement of a two-level system in the Rabi model,
which is coupled to a coplanar-waveguide resonator as a cavity. The response
of the system is calculated using the master equation for the density
operator both numerically in the semi-quantum approximation and analytically
in the semi-classical approximation. Varying the driving amplitude, we
observe two different regimes, analogously to Ref.~\cite{Szombati19}%
: (i)~weak-driving regime and (ii)~strong-driving regime. In the first one,
the weak-driving regime (i), both the ground and excited qubit states are
monitored, displaying their probabilistic energy-level occupations. And when
the power of driving is strong enough, in the strong-driving regime (ii),
the spectral lines converge into a peak centered at the bare cavity and
qubit transition frequencies. We note that the transition between these
regimes can also be referred to as the driven quantum-to-classical
transition \cite{Pietikainen17b, Pietikainen17a, Pietikainen19}.

\section{Model}

We study a two-level system which is coupled to a cavity. The system is
driven by two signals: the high-amplitude driving tone with the frequency $%
\omega _{\mathrm{d}}$ and the low-amplitude probe tone with the frequency $%
\omega $. The qubit-resonator system we consider in the circuit-QED
realization; specifically, it can be the driven transmon-resonator system 
\cite{You07c, Koch07, Bianchetti09, Bishop09}, which is described by the
Jaynes-Cummings Hamiltonian \cite{Schleich} in the two-level approximation: 
\begin{eqnarray}
H &=&\hbar \omega _{\mathrm{r}}a^{\dag }a+\hbar \frac{\omega _{\mathrm{q}}}{2%
}\sigma _{z}+\hbar \mathrm{g}\left( \sigma _{-}a^{\dag }+\sigma _{+}a\right)
+  \label{JC} \\
&&+\hbar \xi \left( a^{\dag }e^{-i\omega t}+ae^{i\omega t}\right) +\hbar
\mathrm{\Omega} \left( \sigma _{+}e^{-i\omega _{\mathrm{d}}t}+\sigma _{-}e^{i\omega _{%
\mathrm{d}}t}\right) .  \notag
\end{eqnarray}%
Here, $\hbar \omega _{\mathrm{q}}$ is the transition energy between qubit states; $\sigma _{i}$ and $\sigma _{\pm }=\left( \sigma _{x}\pm i\sigma
_{y}\right) /2$ are Pauli operators; the resonator has the quantized
fundamental mode with frequency $\omega _{\mathrm{r}}$; $a^{\dag }(a)$ is
the creation (annihilation) operator of a single excitation in the
resonator; the coupling strength between the two-level system and the
resonator is defined by $\hbar \mathrm{g}$; the probe and drive amplitudes
are described by values $\xi $ and $\mathrm{\Omega} $, respectively. Note that the
transmon-resonator coupling constant $\mathrm{g}$ relates to the bare
coupling $\mathrm{g}_{0}$ as $\mathrm{g}=\mathrm{g}_{0}\sqrt{E_{\mathrm{c}%
}/\left\vert \mathrm{\Delta} -E_{\mathrm{c}}\right\vert }$ with the detuning $\mathrm{\Delta}
=\hbar \left( \omega _{\mathrm{q}}-\omega _{\mathrm{r}}\right) $ and the
qubit charging energy $E_{\mathrm{c}}$, where this renormalization is due to
the virtual transitions through the upper transmon states. In the
experiments, e.g. Refs.~\cite{Bishop09, Bianchetti09, Oelsner13},
the measured value is the normalized transmission amplitude $A$. This is
related to the photon field $\left\langle a\right\rangle $ in the cavity 
\cite{Bianchetti09, Bishop09, Macha14, Rakhmanov08}, $A=2V_{0}\left\vert \left\langle
a\right\rangle \right\vert $, where $V_{0}$ is a voltage related to the gain
of the experimental amplification chain \cite{Bishop09} and it is defined as 
\cite{Bianchetti09} $V_{0}^{2}=Z\hbar \omega _{\mathrm{r}}\varkappa /4$ with 
$Z$ standing for the transmission-line impedance.

\section{Semi-classical approach: analytical solution}

To quantitatively describe the system, we first take the master equation in
the form of Eq.~(5) from Ref.~\cite{Bianchetti09}. This includes the
resonator relaxation rate $\kappa $\ and qubit relaxation and pure dephasing
rates, $\Gamma _{1}$ and $\Gamma _{2}=\Gamma _{\phi }+\Gamma _{1}/2$. While
the equations can be numerically solved in their available form, it is more
illustrative to study analytical expressions first. For this purpose, we
consider the steady-state solution in the so-called semi-classical
approximation, where all correlators are assumed to factorize, $%
\!\left\langle a^{\dag }\sigma \right\rangle \approx \!\left\langle a^{\dag
}\right\rangle \!\left\langle \sigma \right\rangle $, etc.~\cite{Shevchenko18}
Then from the Lindblad equation, we obtain the non-linear equations for $%
\left\langle a\right\rangle $, $\left\langle \sigma \right\rangle $, and $%
\left\langle \sigma _{z}\right\rangle $. These can be rewritten for $%
n=\left\vert \left\langle a\right\rangle \right\vert ^{2}$ and $\left\langle
\sigma _{z}\right\rangle =2P_{+}-1$, with $P_{+}$ standing for the qubit
upper-level occupation probability, as follows

\begin{equation}
n=\left\vert \left\langle a\right\rangle \right\vert ^{2}=\frac{\xi ^{2}}{%
\left[ \chi \left( 2P_{+}-1\right) +\delta \omega _{\mathrm{r}}\right]
^{2}+\kappa ^{2}},  \label{n}
\end{equation}%
\begin{equation}
P_{+}=\frac{1}{2}\frac{\mathrm{\Omega} ^{2}}{\mathrm{\Omega} ^{2}+\mathrm{\Omega} _{1}^{2}(n)},
\label{P}
\end{equation}

\begin{equation}
\mathrm{\Omega} _{1}^{2}(n)=\frac{4}{T_{1}T_{2}}+4\frac{T_{2}}{T_{1}}\left( \delta
\omega _{\mathrm{q}}+2\chi n\right) ^{2}.  \label{Wc}
\end{equation}%
Here $\delta \omega _{\mathrm{r}}=\omega _{\mathrm{r}}-\omega $, $\delta
\omega _{\mathrm{q}}=\omega _{\mathrm{q}}-\omega _{\mathrm{d}}$, $\chi =%
\mathrm{g}^{2}/\mathrm{\Delta} $ describes the dispersive shift, and $\mathrm{\Omega} _{1}$
can be interpreted from Eq.~(\ref{P}), as the characteristic driving
amplitude $\mathrm{\Omega} $, at which the excited qubit level becomes significantly
occupied. In particular, at resonance ($\delta \omega _{\mathrm{q}}=0$) in
the linear approximation (neglecting the term with $n^{2}$), we have the
characteristic driving amplitude as in Ref.~\cite{Bishop09}: $\mathrm{\Omega}
_{1}=2\left( T_{1}T_{2}\right) ^{-1/2}$. In general case, the characteristic
driving amplitude $\mathrm{\Omega} _{1}$ depends on the measurement amplitude $\xi $
and is defined by Eq.~(\ref{Wc}) together with Eqs.~(\ref{n}) and~(\ref{P}).

\section{Semi-quantum approach: numerical solution}

\begin{figure}[ht]
\centering
\includegraphics[width=6cm]{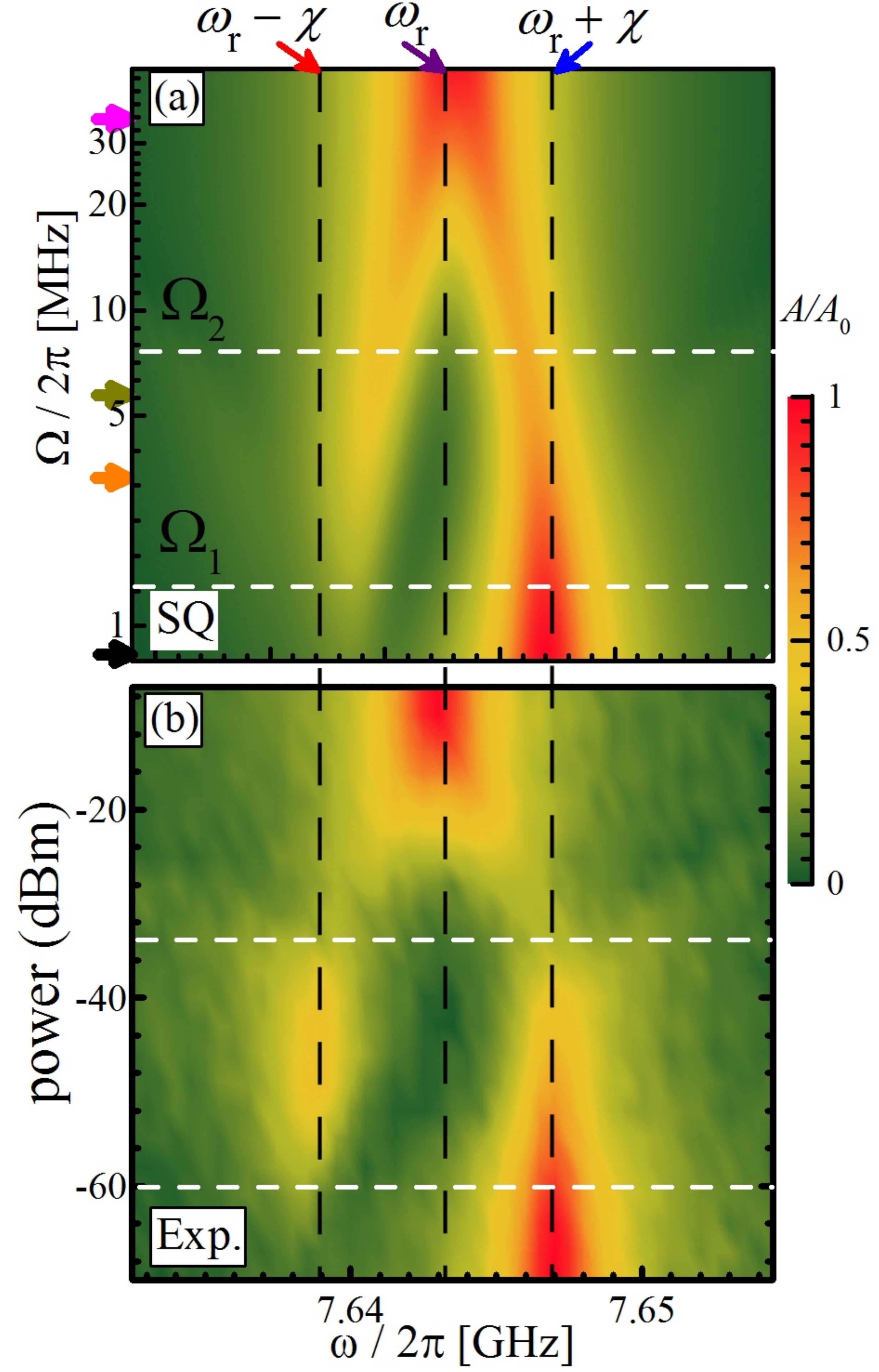}
\caption{Regimes for the driven and probed qubit and characteristic driving
amplitudes $\mathrm{\Omega} _{1}$ and $\mathrm{\Omega} _{2}$. (a) Normalized transmission
amplitude $A$ as a function of the probe frequency $\protect\omega $ and
driving amplitude $\mathrm{\Omega} $. The picture is a result of the steady-state
solution of Maxwell-Bloch equations~(\protect\ref{SQ}). Horizontal colour
arrows show position of the cuts, presented in the next figure. (b)
Respective experimental measurement.}
\label{Fig:2-to-1}
\end{figure}

The system Hamiltonian (\ref{JC}) can be rewritten in a more illustrative
form for large detuning $\mathrm{\Delta} $ between a transmon qubit (in the two-level
approximation) and a resonator mode, $\mathrm{\Delta} \equiv \hbar \left( \omega _{%
\mathrm{q}}-\omega _{\mathrm{r}}\right) \gg \mathrm{g}$,%
\begin{eqnarray}
H^{\prime } &=&\hbar (\delta \omega _{\mathrm{r}}+\chi \sigma _{z})a^{\dag
}a+\hbar \frac{\delta \omega _{\mathrm{q-d}}+\chi }{2}\sigma _{z}+  \notag \\
&&+\hbar \xi (a^{\dagger }+a)+\hbar \frac{\mathrm{\Omega} }{2}(\sigma +\sigma
^{\dagger }),
\end{eqnarray}%
where $\delta \omega _{\mathrm{q-d}}=\omega _{\mathrm{q}}-\omega _{\mathrm{d}%
}$ and $\delta \omega _{\mathrm{q-r}}=\omega _{\mathrm{r}}-\omega $. It is
reasonable to describe the system in the dispersive approximation of the
Jaynes-Cummings Hamiltonian, as in Refs.~%
\cite{Koch07, Bianchetti09,
Oelsner10}. Then, following Refs.~%
\cite{Bianchetti09,
Andre09,Shevchenko18}, we obtain equations in the so-called semi-quantum
model, where one assumes factoring of higher order terms and keeping only
the second order correlators as the following: $\langle a^{\dagger }a\sigma
	_{i}\rangle \thickapprox \langle a^{\dagger }a\rangle \langle \sigma
_{i}\rangle $ and $\langle a^{\dagger }aa\sigma _{i}\rangle \thickapprox
\langle a^{\dagger }a\rangle \langle a\sigma _{i}\rangle $. This allows
truncating the infinite series of equations. From the Lindblad equation, for
the expectation values of the operators $\langle \sigma _{i}\rangle $ $%
(i=x,y,z)$ and the resonator field operators $\left\langle a\right\rangle ,$ 
$\langle a\sigma _{i}\rangle ,$ and $\langle a^{\dagger }a\rangle $ we
obtain the system of equations, known as the Maxwell-Bloch equations:

\begin{subequations}
\label{SQ}
\begin{eqnarray}
\frac{d}{dt}\langle \sigma _{z}\rangle  \!&=&\! \mathrm{\Omega} \langle \sigma
_{y}\rangle -\Gamma _{1}\left( 1+\frac{\langle \sigma _{z}\rangle }{z_{0}}%
\right) , \\
\frac{d}{dt}\langle \sigma _{x}\rangle  \!&=&\! -\!\left( 2\chi
\!\left\langle a^{\dagger }a\right\rangle \!+\!\delta \omega _{\mathrm{q-d}%
}+\!\chi \right) \!\langle \sigma _{y}\rangle -\Gamma _{2}\!\frac{\langle
\sigma _{x}\rangle }{z_{0}}\!, \\
\frac{d}{dt}\langle \sigma _{y}\rangle \! &=&\!\left( 2\chi
\left\langle a^{\dagger }a\right\rangle +\delta \omega _{\mathrm{q-d}}+\chi
\right) \langle \sigma _{x}\rangle - \\
&&-\Gamma _{2}\frac{\langle \sigma _{y}\rangle }{z_{0}}-\mathrm{\Omega} \langle
\sigma _{z}\rangle ,  \notag \\
\frac{d}{dt}\langle a\rangle \! &=&\!-i\left( \delta \omega _{\mathrm{r}}\langle a\rangle +\chi \langle a\sigma _{z}\rangle +\xi \right)
-\kappa \langle a\rangle , \\
\frac{d}{dt}\langle a^{\dagger }a\rangle \! &=&\!-2\xi \text{%
\thinspace Im\thinspace }\langle a\rangle +2\kappa \left( N_{\mathrm{th}%
}-\langle a^{\dagger }a\rangle \right) , \\
\frac{d}{dt}\langle a\sigma _{z}\rangle \! &=&\!-i\left( \delta
\omega _{\mathrm{r}}\left\langle a\sigma _{z}\right\rangle +\chi \langle
a\rangle +\xi \langle \sigma _{z}\rangle \right) + \\
&&+\mathrm{\Omega} \langle a\sigma _{y}\rangle -\Gamma _{1}\left\langle
a\right\rangle -\left( \frac{\Gamma _{1}}{z_{0}}+\kappa \right) \langle
a\sigma _{z}\rangle ,  \notag \\
\frac{d}{dt}\langle a\sigma _{x}\rangle \! &=&\!-i\delta \omega _{%
\mathrm{r}}\left\langle a\sigma _{x}\right\rangle -\left( \frac{\Gamma _{2}}{%
z_{0}}+\kappa \right) \langle a\sigma _{x}\rangle - \\
&&-\left( \delta \omega _{\mathrm{q-d}}+2\chi \left( \langle a^{\dagger
}a\rangle +1\right) \right) \langle a\sigma _{y}\rangle -i\xi \langle \sigma
_{x}\rangle ,  \notag \\
\frac{d}{dt}\langle a\sigma _{y}\rangle \! &=&\!-i\delta \omega _{%
\mathrm{r}}\left\langle a\sigma _{y}\right\rangle -\left( \frac{\Gamma _{2}}{%
z_{0}}+\kappa \right) \langle a\sigma _{y}\rangle - \\
&&-i\xi \langle \sigma _{y}\rangle -\mathrm{\Omega} \langle a\sigma _{z}\rangle + 
\notag \\
&&+\left( \delta \omega _{\mathrm{q-d}}+2\chi \left( \langle a^{\dagger
}a\rangle +1\right) \right) \langle a\sigma _{x}\rangle .\notag
\end{eqnarray}

We have numerically solved the system of equations~(\ref{SQ}) and the
results are presented in Figs.~\ref{Fig:2-to-1}-\ref{Fig4}, where the
normalized transmission amplitude is plotted as a function of the driving
power and probing frequency. For calculations we take the following
parameters, close to the ones of Ref.~\cite{Szombati19}: 
\end{subequations}
\begin{eqnarray}
\frac{\omega _{\mathrm{r}}}{2\pi } &=&7.643~\text{GHz, \ \ }\frac{\omega _{%
\mathrm{q}}}{2\pi }=6.440~\text{GHz, \ \ }\frac{\chi }{2\pi }=4.1~\text{MHz},
\notag \\
T_{1} &=&1.55~\mu \text{s, \ \ }T_{2}=2.65~\mu \text{s, \ \ }\ \kappa /2\pi
=1~\text{MHz.}  \label{parameters}
\end{eqnarray}

\section{Experiment}

Our sample consists of a transmon qubit (named Qubit 2 in Ref.~\cite{Szombati19}) with a transition frequency between its ground and
first excited states $\omega _{\mathrm{q}}/(2\pi )=6.44$~GHz coupled to a
superconducting co-planar waveguide resonator with resonance frequency $%
\omega _{\mathrm{r}}/(2\pi )=7.643$~GHz. The qubit state can be controlled
by applying a coherent drive tone with frequency $\omega _{\mathrm{d}}\simeq
\omega _{\mathrm{q}}$ through a separate charge line. The interaction
between the qubit and the resonator results in a shift of the resonator
frequency dependent on the qubit state. This dispersive shift of $\chi
/(2\pi )=4.1$~MHz allows us to perform the read out of the qubit state by
applying a probe tone to the resonator at frequency $\omega _{\mathrm{p}%
}\simeq \omega _{\mathrm{r}}$ and by subsequent detection of the transmitted
signal with a standard heterodyne detection scheme (see Ref.~%
\cite{Szombati19}).

\section{Results}

\begin{figure}[ht]
\centering
\includegraphics[width=8cm]{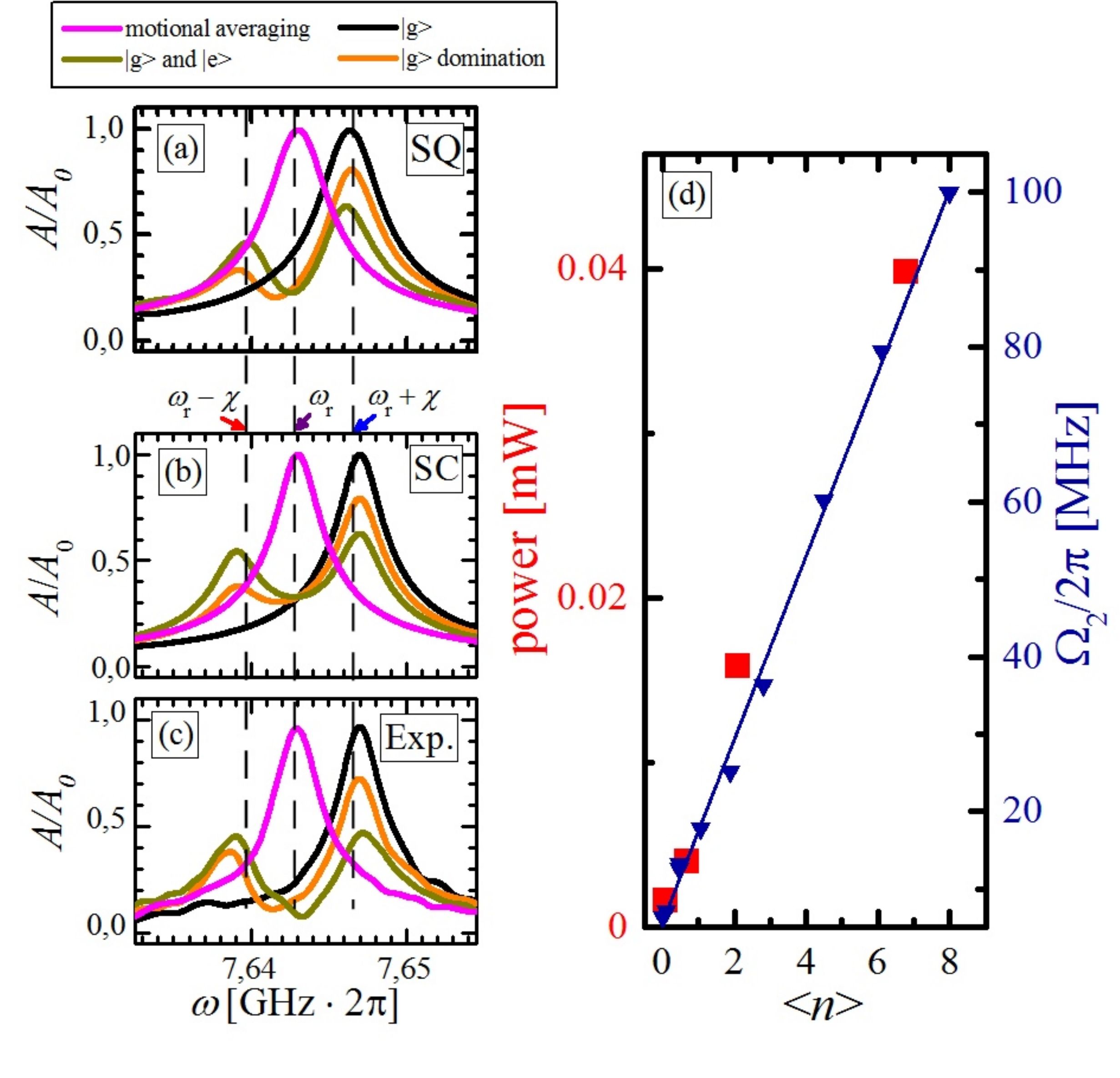}
\caption{Shift of the resonance lines. Numerical semi-quantum (a) and
analytical semi-classical (b) dependencies of the normalized transmission
amplitude $A$ on the probing frequency for several values of driving power $%
\mathrm{\Omega} \,$, shown by the arrows in Fig.~\protect\ref{Fig:2-to-1}. Solid
lines relate to regime (i) with two qubit lines and the dashed lines are for
the regime (ii) with one motional-averaged line. (c) Respective experimental
measurement. (d) Characteristic driving amplitude $\mathrm{\Omega} _{2}$ as a
function of the photon number $\left\langle n\right\rangle $; right scale is
for the theoretical points ($\mathrm{\Omega} _{2}$ in MHz) and left scale is for the
experimental points (power in mW); the agreement between the data here
could be used for the calibration of the power.}
\label{Fig3}
\end{figure}

Numerical and analytical results of the previous two sections together with
the experimental observations are presented in Figs.~\ref{Fig:2-to-1}-\ref%
{Fig4}. Namely, in Fig.~\ref{Fig:2-to-1}(a) we plot the transmission
amplitude $A$, normalized to its maximal value $A_{0}$, as a function of the
probe frequency $\omega $ and driving amplitude $\mathrm{\Omega} $, which is
calculated using the equations and parameters from the previous section,
Eqs.~(\ref{SQ}-\ref{parameters}). Several horizontal cuts of Fig.~\ref%
{Fig:2-to-1}(a) are presented as Fig.~\ref{Fig:2-to-1}(a-c), which is the
transmission plotted as a function of the probe frequency $\omega $ for
several values of the drive amplitude $\mathrm{\Omega} $ (see explanations below).

In Fig.~\ref{Fig:2-to-1}(a) we can observe two different regimes. Let us now
analyze these regimes in more detail, since these present the main result of
our work here.

(i) $\mathrm{\Omega} \ll \mathrm{\Omega} _{2}$. This can be called the \textquotedblleft
fast-measurements\textquotedblright\ regime, or equivalently weak-driving
regime, or \textquotedblleft quantum\textquotedblright\ regime. This is
because the two qubit states are visualized with the position of the
respective resonances at $\delta \omega =\pm \chi $. So, in this regime both
ground and excited qubit states are monitored, with respective
probabilities. Namely, with increasing the driving amplitude $\mathrm{\Omega} $, the
probability of finding the qubit in the excited state increases, left line
in Fig.~\ref{Fig:2-to-1}(a), while the probability\textbf{} of the ground state
(right resonance line) decreases. See also about this, e.g., in Ref.~%
\cite{Reagor16}.

(ii) $\mathrm{\Omega} \gg \mathrm{\Omega} _{2}$. This can be called the \textquotedblleft
slow-measurements\textquotedblright\ regime, or equivalently strong-driving
regime, or \textquotedblleft classical\textquotedblright\ regime. This
corresponds to no frequency shift, with qubit states equally populated,
which thus can also be referred to as a \textquotedblleft
motional averaging\textquotedblright . Similar transitions from a \textquotedblleft
quantum\textquotedblright\ to \textquotedblleft classical\textquotedblright\
regime was observed in Refs.~%
\cite{Fink10, Li13, Pietikainen17a, 
Pietikainen17b, Szombati19}.

Figure~\ref{Fig3}(d) shows the dependence of the characteristic driving
amplitude $\mathrm{\Omega} _{2}$, which separates regimes (i) and (ii), as a function
of photon numbers obtained from our numerical solution as well as the
experimental points. The theoretical points were calculated for the probe
amplitudes $\xi $ from $0.01\kappa $ to $2\kappa $. The experimental points
were taken from Fig.~S4 of Ref.~\cite{Szombati19}.

Let us now consider interpretation of the above results analytically, for
the two respective regimes.

Regime (i). This is not described directly by the stationary solution. To
describe this regime we note that in this case the qubit is found either in
the ground state with the probability $P_{-}$ or in the excited state with
the probability $P_{+}$. Then the measured normalized transmission amplitude
can be calculated as following 
\begin{equation}
A=P_{-}A_{-}+P_{+}A_{+}  \label{PAPA}
\end{equation}%
with the partial values $A_{\pm }$ given by Eq.~(\ref{n}):%
\begin{equation}
A_{\pm }=\frac{\xi }{\sqrt{\left[ \pm \chi +\delta \omega _{\mathrm{r}}%
\right] ^{2}+\kappa ^{2}}}.  \label{Apm}
\end{equation}%
With these formulas we plot the solid lines in Fig.~\ref{Fig3}(b).

Regime (ii). Under the strong resonant driving, the qubit levels are equally
populated and then Eq.~(\ref{n}) with $P_{+}=1/2$ yields%
\begin{equation}
A=\frac{\xi }{\sqrt{\delta \omega _{\mathrm{r}}^{2}+\kappa ^{2}}}\text{.}
\end{equation}%
With this formula we plot the magenta solid line in Fig.~\ref{Fig3}(a).

\begin{figure}[ht]
\centering
\includegraphics[width=\columnwidth]{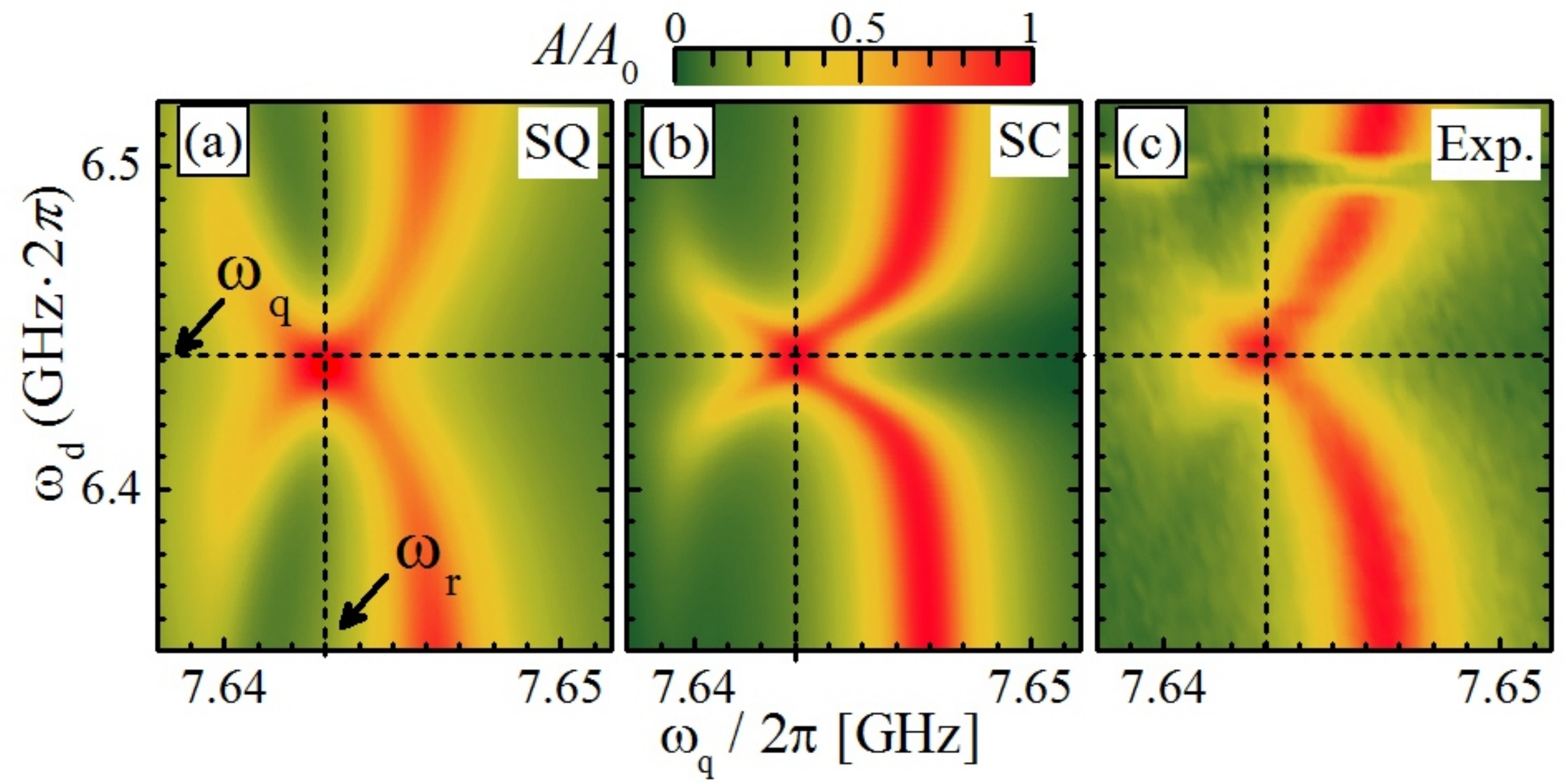}
\caption{The resonant transmission shift. Analytical (a) and semi-quantum
(b) dependencies of the normalized transmission amplitude $A$ on the probe
frequency $\protect\omega $ and the drive frequency $\protect\omega _{%
\mathrm{d}}$. (c) Respective experimental measurement.}
\label{Fig4}
\end{figure}

Finally we describe both the qubit and resonator frequency shifts. For this,
we make use of Eq.~(\ref{PAPA}) with the partial transmission amplitudes $%
A_{\pm }$ defined by the respective probabilities, rather than assuming them
equal to $0$ or $1$ as in Eq.~(\ref{Apm}), as following%
\begin{equation}
A_{\pm }=\frac{\xi }{\sqrt{\left[ \chi \left( 1-2P_{\pm }\right) +\delta
\omega _{\mathrm{r}}\right] ^{2}+\kappa ^{2}}}.  \label{Apm2}
\end{equation}%
With this analytical formulas we plot Fig.~\ref{Fig4}(b), which is
remarkably in agreement with the numerical result in Fig.~\ref{Fig4}(a) and
the experiment in Fig.~\ref{Fig4}(c).

Note that the advantage of the analytical results, presented here, is that
these are transparent formulas, which capture the main physics here.
Importantly, these results are confirmed by the numerical calculations, done
within the semi-quantum model, and by comparison with the experiments.

\section{Conclusions}

We have studied the interaction between a cavity and a two-level system in
the Jaynes-Cummings model with dispersive coupling under external drive. We
have demonstrated that there are two different regimes. The quantum and
semi-classical regimes demonstrate detection of the two energy transitions
in the system, while at higher driving amplitude the two resonance lines
merge into one, making a specific motional-averaging picture, defined by the
qubit energy-level occupation probabilities. We believe that our analytical
and numerical results are useful for deeper understanding of both
experimental realizations and theoretical description of dynamical phenomena
in circuit quantum electrodynamics.

\begin{acknowledgements}
	D.S.K. is grateful to A.~Sultanov for fruitful discussions. D.S.K.
acknowledges the hospitality of the Leibniz Institute of Photonic
Technology, where part of this work was done. The work of D.S.K. was partly
supported by DAAD Binationally Supervised Doctoral Degrees Research Grant
(No.~57299293) and DAAD Short-Term Research Grant (No.~57381332). S.N.S. and
D.S.K. acknowledge partial support by the Grant of the President of Ukraine (Grant No.~F84/185-2019).
D.S., A.G.F. and A.F. were supported by the Australian Research Council
Centre of Excellence for Engineered Quantum Systems (EQUS, CE170100009).

	All the authors were involved in the preparation of the
manuscript. All the authors have read and approved the final manuscript.
\end{acknowledgements}

\nocite{apsrev41Control} 
\bibliographystyle{apsrev4-1}
\bibliography{2_to_1_bibliography}

\end{document}